# Importance of quantum interference in molecular-scale devices

Kamil Walczak [1]

Institute of Physics, Adam Mickiewicz University
Umultowska 85, 61-614 Poznań, Poland

Electron transport is theoretically investigated in a molecular device made of anthracene molecule attached to the electrodes by thiol end groups in two different configurations (para and meta, respectively). Molecular system is described by a simple Hückel-like model (with non-orthogonal basis set of atomic orbitals), while the coupling to the electrodes is treated through the use of Newns-Anderson chemisorption theory (constant density of states within energy bandwidth). Transport characteristics (current-voltage and conductance-voltage) are calculated from the transmission function in the standard Landauer formulation. The essential question of quantum interference is discussed in detail. The results have shown a striking variation of transport properties of the device depending on the character of molecular binding to the electrodes.

Key words: molecular device, quantum interference, electronic transport, molecular electronics

PACS numbers: 85.65.+h , 73.23.-b

## I. Introduction

Molecular junctions are promising candidates as future electronic devices because of their small size and self-assembly features. Such junctions are usually composed of two metallic electrodes (source and drain) joined by individual molecule (bridge). The charge is transferred under the bias voltage and current-voltage (I-V) characteristics are measured experimentally [1]. In general, transport properties of such structures are dominated by some effects of quantum origin, such as: tunneling, quantization of molecular energy levels and discreteness of electron charge and spin. However, recently it was pointed out that also quantum interference effects can lead to substantial variation in the conductance of molecule-scale devices [2-9].

The main purpose of this work is to show some theoretical aspects of interference phenomena in anthracene molecule connecting two identical electrodes by thiol (–SH) end groups (see fig.1). These end groups (or more precisely sulfur terminal atoms, since hydrogen atom seems to be lost in the chemisorption process) ensure readily attachment to metal surfaces [10]. It is shown that the molecule acts not only as a scattering impurity between two reservoirs of electrons (electrodes), but simultaneously as an "electronic interferometer". Interference itself reveals the wave nature of the electrons passing from the source to drain through the molecule. Here the variation of interference conditions is achieved by changing the connection between anthracene molecule and electrodes.

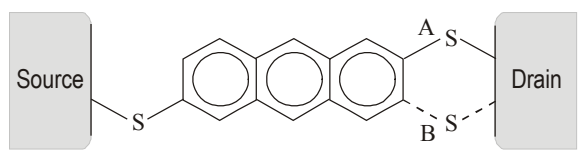

Fig.1 A schematic model of analyzed samples.

## II. Theoretical treatment

Molecular device is defined as anthracene molecule joined to two metallic surfaces with the help of thiol end groups in two different configurations – para (A) and meta (B), respectively. In both cases we have different interference conditions and so we expect to observe changes in transport characteristics. Problem of electronic conduction between two continuum reservoirs of states via a molecular bridge with discrete energy levels can be solved within transfer matrix technique of scattering theory [11,12]. The current flowing through the device is obtained from the transmission function T through the integration procedure [12]:

$$I(V) = \frac{2e}{h} \int_{-\infty}^{+\infty} T(E)[f(E - \mu_S) - f(E - \mu_D)]dE, \qquad (1)$$

where: f denotes Fermi distribution function for room temperature (293 K) with chemical potentials $\mu_{S/D} = E_F \pm eV/2$ referred to the source and drain, respectively. In this type of non-self-consistent calculations, one must postulate voltage distribution along the molecular bridge. For the sake of simplicity we assume that voltage drop is limited to the electrodes only [13], shifting their Fermi level located in the middle of the HOMO-LUMO gap [14]. However, other choices of the voltage distribution have only a small effect on our final results and general conclusions. The differential conductance is then calculated as the derivative of the current with respect to the voltage [15]:

$$G = \tfrac{1}{2} G_0 [T(\mu_S) + T(\mu_D)], \qquad (2)$$

where $G_0 = 2e^2/h \approx 77.5$ [µS] is the quantum of conductance.

Formula for the transmission probability can be expressed in the convenient matrix form [12]:

$$T(E) = tr[(\Sigma_S - \Sigma_S^+)G(\Sigma_D^+ - \Sigma_D)G^+], \qquad (3)$$

where $\Sigma_{S/D}$ and are self-energy terms of the source/drain electrode and the Green's function of the molecule is expressed as follows:

$$G = [ES - H - \Sigma_S - \Sigma_D]^{-1}. \qquad (4)$$

Here $S$ denotes overlapping matrix (where the overlap between the nearest-neighbor sites is assumed to be equal to 0.25). Since only delocalized π-electrons dictate the transport properties of organic molecules, the electronic structure of the molecule is described by a simple Hückel Hamiltonian H with one π-orbital per site (atom) [16], where overlapping is explicitly included (using non-orthogonal basis set of atomic orbitals). Throughout this work we take the standard energy parameters for organic conjugated systems: on-site energy is $\alpha = -6.6$ eV and nearest-neighbor hopping integral is $\beta = -2.7$ eV. In the Hückel π-bond picture, all carbon and sulfur atoms are treated equivalently (because of their electronegativity). In our simplified model, the coupling to the electrodes is treated through the use of Newns-Anderson chemisorption theory [11], where ideal electrodes are described by constant density of states within energy bandwidth [17-20]. So self-energy matrices ($\Sigma$) take the diagonal form with elements equal to $-0.05i$ [eV].

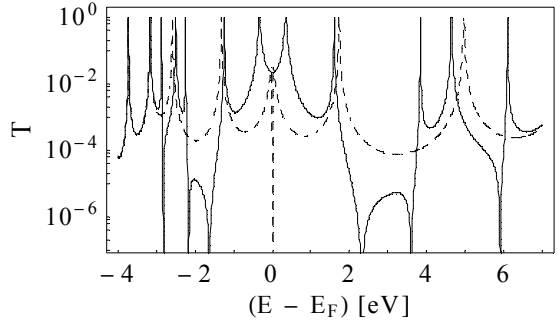

Fig.2 Transmission as a function of electron energy (with respect to Fermi energy level) for devices in configuration A (solid curve) and B (broken curve), respectively.

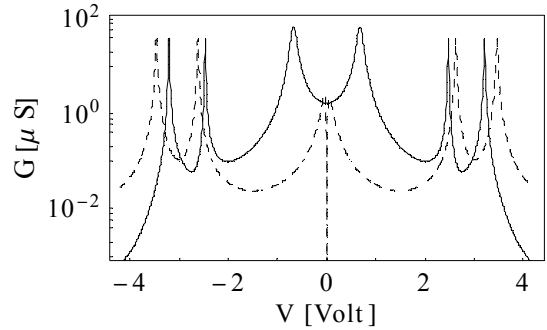

Fig.3 Comparison of conductance spectra for devices in configuration A (solid curve) and B (broken curve), respectively.

## III. Results and discussion

Now we proceed to analyze our results from the point of view of quantum interference effects. The geometry of the molecule is taken to be that of anthracene with sulfur atoms on either end of the molecule, binding it to the electrodes in two different configurations – para (A) and meta (B), respectively. For isolated anthracene the HOMO is at $-7.614$ eV and the LUMO is at $-5.352$ eV. Because of our simplification that Fermi level is arbitrarily chosen to be located in the middle of the HOMO-LUMO gap, $E_F = -6.483$ eV. The HOMO-LUMO gap for molecular system in para configuration is reduced from 2.262 eV for anthracene to the value of 0.667 eV, but for molecular system in meta configuration it is reduced to zero.

Figure 2 shows the transmission dependence on the electron energy for anthracene in para (A) and meta (B) connections with identical electrodes. For transparency we plot it in the logarithmic scale. Asymmetry of the transmission function (with respect to the Fermi energy level) is due to non-orthogonality of atomic orbitals used to describe molecular system. The existence of resonances in the transmission probability is associated with resonant tunneling through molecular eigenstates. Such resonance peaks are shifted and broadened by the fact of the coupling with the electrodes (just like discrete energy levels of the molecule). A change in the configuration of connection between anthracene and two electrodes results in variation of the interference conditions and obvious changes in the transmission function. It manifests itself as shifts in the resonance peaks and in reduction of their height. Well-separated energy levels give rise to distinct peaks in the spectrum, while molecular levels close in energy can overlap and eventually interfere (reduction of resonance peaks is due to destructive interference).

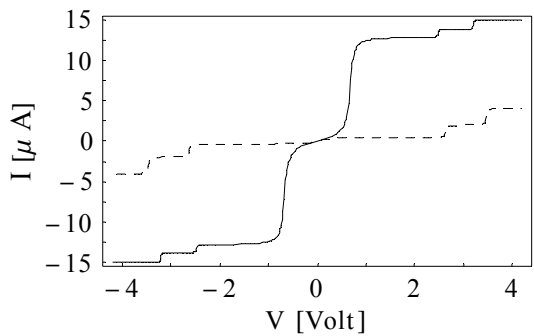

Fig.4 Comparison of current-voltage characteristics for devices in configuration
A (solid curve) and B (broken curve), respectively.

Another remarkable feature of the transmission spectrum is the appearance of antiresonances, which are defined as transmittance zeros and correspond to the physical situation for incident electron being perfectly reflected by a molecule. There are two different mechanisms (well-known in literature) responsible for the origin of antiresonances. One of these is associated with interference between the different molecular orbitals through which the electron propagates [2,21]. The second mechanism is due entirely to the non-orthogonality of atomic orbitals on different atoms [17]. In principle, transport problem in which a non-orthogonal basis set of states is used can be solved by a method proposed recently by Emberly and Kirczenow [5], where condition for antiresonances was analytically demonstrated. However, in this work we perform numerical evaluations of energies at which incoming electron has no chance to leave the source electrode. There are six antiresonances for device in configuration A ($E = -2.821 + E_F$, $-2.160 + E_F$, $-1.622 + E_F$, $2.320 + E_F$, $3.600 + E_F$, $5.907 + E_F$) and only one for device in configuration B ($E = E_F$). Antiresonance is predicted to manifest itself by producing a drop in the differential conductance [5]. Moreover, the fact that it is generated exactly at the Fermi energy of metallic electrodes has important consequences for the conductance spectrum in which antiresonance can be observed (as shown in fig.3). However, in practice this unusual phenomenon can be blurred by some neglected factors which are present in realistic systems, such as: Stark effect, σ states, σ-π hybridization or many-body effects.

In figure 4 we plot the current-voltage (I-V) characteristics for both analyzed structures (in para – A and meta – B connections, respectively). The current steps are attributed to the discreteness of molecular energy levels as modified by the coupling with the electrodes [12]. Because this coupling is assumed to be small (bad contacts are suggested by experimental data [1]), the transmission peaks are very narrow and therefore the I-V dependence has a step-like character. In particular, the height of the step in the I-V curve is directly proportional to the area of the corresponding peak in the transmission spectrum. Since quantum interference is important in determining the magnitudes of the resonance peaks, it is also crucial for calculations of the tunneling current. Indeed, the magnitude of the current flowing through the device is very sensitive on the manner of attachment between anthracene molecule and metal surfaces. Large values of the current are predicted for device of configuration A, while reduction of the current by orders of magnitude is expected for device of configuration B (although the shape of the I-V curve is similar in both cases). Such reduction is caused by destructive interference.

## IV. Summary

In this paper we have examined the possibility that quantum interference can substantially affect the conductance in molecular-scale devices. The results have shown a striking variation of all the transport characteristics depending on the geometry of the molecular system (its connection with the electrodes). Anyway, the quantum effect of destructive interference may be used within the molecular device to switch its conductivity on and off [8,9]. The existence of interference effects in molecular devices open the question of their control. The phase shift of molecular orbitals could be controlled by a transverse magnetic field or a longitudinal electric field. However, magnetic field seems to be too large to produce significant phase shift (according to our simulations – hundreds of Teslas).

## Acknowledgments

Author is very grateful to B. Bułka, T. Kostyrko and B. Tobijaszewska for illuminating discussions. Special thanks are addressed to S. Robaszkiewicz for his stimulating suggestions.